\documentclass[11pt]{article}
\usepackage{amsmath,amssymb,color,graphics,epsfig}
\usepackage{amsfonts}
\newcommand{\hoch}[1]{$\, ^{#1}$}

\newcommand{\be}{\begin{equation}}
\newcommand{\ee}{\end{equation}}
\newcommand{\bea}{\setlength\arraycolsep{2pt} \begin{eqnarray}}
\newcommand{\eea}{\end{eqnarray}}
\newcommand{\nn}{\nonumber}

\def\ft#1#2{{\textstyle{\frac{\scriptstyle #1}{\scriptstyle #2} } }}
\def\fft#1#2{{\frac{#1}{#2}}}

\def\0{{\sst{(0)}}}
\def\1{{\sst{(1)}}}
\def\2{{\sst{(2)}}}
\def\3{{\sst{(3)}}}
\def\4{{\sst{(4)}}}
\def\5{{\sst{(5)}}}
\def\6{{\sst{(6)}}}
\def\7{{\sst{(7)}}}
\def\8{{\sst{(8)}}}
\def\sst#1{{\scriptscriptstyle #1}}

\begin{document}

\begin{flushright}
%\hfill{KIAS-P12028}
 %\hfill{
%\bf hep-th/yymmnnn}
\end{flushright}

\vspace{25pt}
\begin{center}
{\large {\bf Isotropic Expansion of Inhomogeneous Universe}}

\vspace{10pt}
Wei-Jian Geng and H. L\"u

\vspace{10pt}

{\it \hoch{1}Department of Physics,
Beijing Normal University, Beijing 100875, China}

\vspace{40pt}

\underline{ABSTRACT}
\end{center}

We propose a cosmological model that describes isotropic expansion of inhomogeneous universe. The energy-momentum tensor that creates the spatial inhomogeneity may not affect the uniform expansion scaling factor $a(t)$ in the FLRW-like metrics.  Such energy-momentum tensor may not be exotic; in fact any linear or non-linear $\sigma$-model has this feature. We show that the classical spatial inhomogeneity can be embedded in both inflation models and the traditional cosmological expansion by perfect fluid.  The spatial inhomogeneity resembles the primordial quantum perturbation that was frozen in the co-moving frame. We obtain some exact inhomogeneous solutions with spherical or axial symmetries.  We also show that some of our cosmological models can be viewed as the dynamical black hole formation.

\vfill {\footnotesize Emails: gengwj@mail.bnu.edu.cn\ \ \ mrhonglu@gmail.com}

\thispagestyle{empty}

\pagebreak
%\voffset=0pt
%\setcounter{page}{1}

%\tableofcontents
%\addtocontents{toc}{\protect\setcounter{tocdepth}{2}}

%%%%%%%%%%%%%%%%%%%%%%%%%%%%%%%%%%%%%%%%

\newpage
%%%%%%%%%%%%%%%%%%%%%%%%%%%%%%%%%%%%%%%%

{\bf Introduction:} The recent BICEP observation \cite{Ade:2014xna}, if confirmed, may provide strong evidence for the inflationary origin \cite{Guth:1980zm,Linde:1981mu} of our universe.  In the discussion of the cosmological expansion of the universe, aside from the most general solutions, there can exist three simplified scenarios for our 3-space: (1) isotropic and homogeneous; (2) homogeneous but anisotropic and (3) isotropic but inhomogeneous.  The first case is captured by the FLRW model, which can also be used to describe the cosmological inflation when certain inflaton is present.  This model is consistent with our observation of the universe at large scale.  The small causally-disconnected inhomogeneity can then be explained by the early quantum fluctuation that was frozen by the inflation.  One can also entertain the possibility of anisotropic universe using the Bianchi IX metrics in which the space is homogeneous, (e.g.~the $SU(2)$ group manifold \cite{Belinskii:1972sg} or the flat 3-space \cite{Watanabe:2009ct},) but with anisotropic expansion.

   The last scenario, isotropic expansion of inhomogeneous universe, has been typically discarded.  Such scenario appears not to satisfy the Einstein's equations of motion. Naively, one would expect that any energy-momentum tensor that could cause the inhomogeneity in the space should affect the expansion rate as well, giving rise to anisotropic expansion in the three spatial directions. However, the inhomogeneity of our universe may be related to the dipole anisotropy found by Planck. (See, e.g.~\cite{Cai:2013gma}.) Furthermore, inhomogeneous models can explain the SN-data \cite{snia}. In this paper, we present a class of models that give rise to isotropic expansion of inhomogeneous universe.

{\bf The theory}:  We consider Einstein gravity minimally coupled two classes of scalar fields $(\Phi, \phi^I)$ in general dimensions:
\be
e^{-1}{\cal L}_D = R - \ft12(\partial\Phi)^2 -V(\Phi) -\ft12 g^{\mu\nu}\partial_{\mu} \phi^{I} \partial_\nu \phi^J h_{IJ}\,.\label{genlag}
\ee
Here the scalar $\Phi$ is treated as inflaton with a scalar potential $V$; the scalars $\phi^I$ can be any linear or non-linear $\sigma$-model, with $h_{IJ}(\phi^K)$ being the metric. The $\phi^I$ fields involve only the kinetic terms with no scalar potential. The equations of motion are
\bea
R_{\mu\nu} &=& \ft12\partial_\mu \Phi \partial_\nu \Phi + \ft12 h_{IJ} \partial_\mu \phi^I \partial_\nu \phi^J + \fft{V}{D-2} g_{\mu\nu}\,,\cr
\Box\Phi &=& \fft{\partial V}{\partial\Phi}\,,\qquad
\Box \phi^I + g^{\mu\nu} \partial_\mu \phi^J \partial_\nu \phi^K
\Gamma^I{}_{JK} =0\,,
\eea
where $\Gamma$ is the affine connection for $h_{IJ}$.

{\bf Cosmological ansatz}:  We consider the cosmological model
\bea
ds_{D}^2 &=& - dt^2 + a(t)^2 ds^2_{D-1}\,,\qquad ds_{D-1}^2 =\tilde g_{ij} dx^i dx^j\,,\label{ansatz0}\\
\Phi &=& \Phi(t)\,,\qquad \phi^I=\phi^I(x^i)\,.\label{ansatz1}
\eea
The metric ansatz is analogous to the FLRW model. A crucial difference is that the principal orbits $ds_{D-1}^2$ in our models are no longer homogeneous such as $S^3$, $T^3$ or $H^3$.  Instead they are given by some inhomogeneous spaces whose inhomogeneity is caused by the scalars $\phi^I$ at the classical level. Our space-time configuration is also different from those spherically-symmetric solutions in which the isotropy extends only in the sphere directions, but not in the radial direction (see, e.g. \cite{Bhattacharya:2009bz}.)

We now demonstrate the ans\"atze (\ref{ansatz0}) and (\ref{ansatz1}) are consistent with equations of motion. It follows from the Ricci curvature of the metric (\ref{ansatz1})
\be
R_{tt} = -\fft{(D-1)\ddot a}{a}\,,\qquad
R_{ij} = \big((D-2)\dot a^2 + a \ddot a\big) \tilde g_{ij} + \tilde R_{ij}\,,
\ee
that the equations of motion become separated into two sets. One involves $(a(t),\Phi(t))$ and the other involves $(\tilde g_{ij},\phi^I)$, which depend on the spatial coordinates only.

{\bf Separation of variables:} The time-dependent set is given by
\bea
&&H^2 = \fft{1}{(D-1)(D-2)} (\ft12 \dot \Phi^2 + V) - \fft{k}{a^2}\,,\qquad
\dot H = - \fft{1}{D-2} (\ft12 \dot \Phi^2) + \fft{k}{a^2}\,,\cr
&&\ddot \Phi + (D-1) H \dot \Phi + \fft{\partial V}{\partial \Phi} =0\,.\label{time-eom}
\eea
where $H=\dot a/a$ is the Hubble parameter.  The scaling symmetry implies that the parameter $k$, introduced from the separation of variables, can take discrete values, $(-1,0,+1)$.  The time-dependent equations (\ref{time-eom}) become precisely those of the (isotropic and homogeneous) FLRW model with matter field $\Phi$, namely
\bea
e^{-1}{\cal L}_D = R - \ft12(\partial\Phi)^2 -V(\Phi)\,,\qquad
ds_D^2 = -dt^2 + a(t)^2 d\Omega_{D-1,k}^2\,.
\eea
The space-dependent equations are
\be
\tilde R_{ij}= \ft12\partial_i \phi^I \partial_j \phi^J h_{IJ} + k \tilde g_{ij}\,,\qquad
\widetilde \Box\phi^I+ \tilde g^{ij}\partial_i \phi^J \partial_j \phi^K
\Gamma^I{}_{JK}=0\,.
\ee
These are of Einstein gravity coupled to the sigmal-model in $(D-1)$-dimensional space: $e^{-1} {\cal L}_{D-1} = \widetilde R -(D-2)k - \ft12 \tilde g^{ij}\partial_{i} \phi^{I} \partial_j \phi^J h_{IJ}$.  Thus our ansatz can be viewed as a warped time-like Kaluza-Klein reduction. (For $k=0$, the Euclidean theory can also be obtained from the reduction on the stationary time; see, e.g.~\cite{gibbons}.)

In the above model, we need at least two scalars: one is the inflaton that depends on time, and the other is time-independent.  It is very important that the time-independent scalar should not have any scalar potential so as not to affect the expansion rate.  Such a scalar can arise as modulus parameter in the flat direction of the string compactification.

{\bf Generalizing the energy-momentum tensor:} The $\sigma$-model fields can be replaced by any matter energy-momentum tensor $(T^{\rm s})_{\mu\nu}$ that satisfies
\be
(T^{\rm s})_{0i}=0\,,\qquad (T^{\rm s})_{00} - \ft12 g_{00} (T^{\rm s})^\mu {}_\mu=0\,,\qquad
\partial_0 \big((T^{\rm s})_{ij} - \ft12 g_{ij} (T^{\rm s})^\mu{}_\mu\big)=0\,.\label{menergy}
\ee
The curvature tensor in $(D-1)$-space then satisfies
\be
\tilde R_{ij} - \ft12 \tilde g_{ij} \big(\tilde R + (D-3)k\big)= T^{\rm s}_{ij}\,.
\ee
The scalar $\Phi(t)$ can be replaced by multi-scalars.  It can also be replaced by some generic perfect fluid with $T^{\rm t}_{\mu\nu} = {\rm diag} \{-\rho(t), p(t)\,\cdots, p(t)\}$.  The general equations of motion for our model is then given by
\be
R_{\mu\nu} +\fft{\Lambda}{D-2} g_{\mu\nu} = T^{\rm t}_{\mu\nu} + T^{\rm s}_{\mu\nu}\,.
\ee
The time-dependent part of equations of motion for the cosmological ansatz (\ref{ansatz0}) are then exactly the same as those of the standard FLRW model.

{\bf Some exact solutions:}  The FLRW equations have been studied extensively in literature. We focus on the equations of the space sector and construct some exact solutions.  We restrict our discussion in 3-space dimensions. We first consider spherically-symmetric solutions in three dimensions:
\be
ds^2_3 = dr^2 + b(r)^2 d\Omega_2^2\,.
\ee
The solution for $b$ is independent of the scalars $\phi^I$, and it is given by
\be
b^2=\left\{
    \begin{array}{ll}
      r^2 - \ell^2, &\qquad\qquad k=0; \\
      A \cosh(2r+ \delta)-\ft12, &\qquad\qquad k=-1; \\
      A \cos(2r)-\ft12, &\qquad\qquad k=+1.
    \end{array}
  \right.
\ee
The scalars can be solved straightforwardly.  For example, for a single free scalar $\phi$ in the $k=0$ case, we have $\phi={\rm arctanh}(r/\ell)$.  It is of interest to note that if we let $\ell\rightarrow {\rm i}\ell$, for which $\phi$ becomes pure imaginary, the metric describes a wormhole of radius $\ell$.  Combining with the time direction, we obtain a wormhole solution in an isotropic expanding universe.

We also obtain an axial-symmetric solution for Einstein gravity coupled to the scalar $SL(2,R)/U(1)$ coset:
$
e^{-1}{\cal L} = R - \ft12 (\partial\phi)^2 - \ft12 e^{2\phi} (\partial\chi)^2\,.
$
We find
\bea
ds_3^2 &=&\big( x(x+q) +p^2\cos^2\theta\big) \Big(\fft{dx^2}{x(x+q)+p^2} +
d\theta^2\Big) + \big(x(x+q)+p^2\big) \sin^2\theta\,d\phi^2\,,\cr
e^{\phi} &=& 1 + \fft{q(x+q)}{ x(x+q) + p^2\cos^2\theta}\,,\qquad \chi = \fft{pq \cos\theta }{(x+q)^2 + p^2\cos^2\theta}\,,
\eea
where $(p,q)$ are two integration constants.

{\bf Cosmology as black hole formation:}  When the inhomogeneous 3-space is spherically symmetric, the four-dimensional cosmological solution is of cohomogeneity-two and the Killing vectors lie in the $S^2$ only.  This is exactly the same situation of black hole formation while keeping the spherical symmetry intact. It is then natural to expect that the cosmological solution of expanding universe may be viewed as black hole formation by a different observer.  To see this, we define a new coordinate $u$ such that $du=dr + dt/a(t)$, we have
\be
ds_4^2= 2a^2 du dr -a^2 du^2 + a^2 b^2 d\Omega_2^2\,.\nn
\ee
Let $R=ab$, the existence of an apparent horizon is then given by
\be
g^{\mu\nu} \partial_\mu \partial_\nu R = -b\fft{d^2a }{dt^2} + a^{-1} \fft{d^2b}{dr^2}=0\,.\nn
\ee

To make the above observation concrete, we consider cosmological Einstein gravity coupled to a free scalar, i.e. $e^{-1} {\cal L} = R - 2\Lambda - \ft12 (\partial \phi)^2$.  Following the earlier discussion, it is easy to construct the cosmological solution
\bea
ds^2 &=& -dt^2 + \fft{\sin^2(\sqrt{\lambda} t)}{\lambda} \Big(dr^2 + \ft14 e^{-2r} \big((e^{2r}-1)^2 - q^2\big) d\Omega_2^2\Big)\,,\cr
e^{\phi} &=& \fft{e^{2r}-1-q}{e^{2r}-1+q}\,,\qquad \lambda = - \ft13 \Lambda\,.
\eea
For positive cosmological constant $\Lambda$, the sine function becomes sinh and the metric describes an expanding universe. Making the following coordinate transformation
\be
\tilde r = -\ft12 e^{-r} (e^{2r}-1 + q)\, \fft{\sin(\sqrt{\lambda}\,t)}{\sqrt{\lambda}}\,,\qquad
v = e^{-r}\, \fft{\sin(\sqrt{\lambda}\,t)}{\sqrt{\lambda}}\,,
\ee
the solution becomes
\be
ds^2 = \fft{2dv d\tilde r + (q-1) dv^2 - \lambda (\tilde r dv - v d\tilde r)^2 }{1 + \lambda v (2\tilde r + (q-1) v)} + \tilde r(\tilde r+q v) d\Omega_2^2\,,\qquad
e^{\phi} = 1 + \fft{q v}{\tilde r}\,.
\ee
For $\lambda=0$, the solution reduces to the Roberts solution \cite{Roberts89} in Eddington-Finkelstein coordinates.  For general $\lambda$, we can define the Eddington-Finkelstein coordinates as follows
\be
v=\fft{2(\lambda \tilde r + w)}{\lambda(1-q) + w^2}\,,\qquad \hbox{with}\qquad w(u)=\fft{\sqrt{\lambda(q-1)}}{\tanh(\ft12\sqrt{\lambda(q-1)}\,u)}\,.
\ee
The metric becomes
\be
ds^2= 2 d\tilde r du - (\lambda \tilde r^2 +1-q) du^2 + \tilde r(\tilde r+q v) d\Omega^2\,.
\ee
It is easy to very that for large $R=\sqrt{\tilde r(\tilde r+q v)}$, the metric is asymptotic to the (A)dS or flat space-times depending on $\Lambda$.  The apparent horizon emerges only for $q>1$, independent of $\Lambda$.  The above procedure allows us to obtain the new exact black hole formation in asymptotic (A)dS from the simpler inhomogeneous cosmological solution.  These solutions do not evolve to some static black holes, unlike those recently found in \cite{Zhang:2014sta}.

{\bf The (A)dS/CFT correspondence:}  The Lagrangian (\ref{genlag}) is also common in gauged supergravities in which the potential $V(\Phi)$ has stationary point giving rise to negative cosmology constant.  In these theories we can construct asymptotic locally AdS geometries
\be
ds_D^2 = \fft{dr^2}{r^2} + a(r)^2 \tilde g_{\mu\nu} dx^\mu dx^\nu\,,\qquad
\Phi=\Phi(r)\,,\qquad \phi^I=\phi^I (x^\mu)\,.
\ee
Asymptotically at $r\rightarrow \infty$, we have $a\sim r$. In general, the behaviour of $\phi^I$ at large $r$ is governed by the boundary condition $\phi^I \sim \phi_1^I(x) + \phi_2^I(x)/r^{D-1}$. In the AdS/CFT correspondence \cite{adscft}, $\phi_1^I$'s are treated as the sources and $\phi_2^I$'s are the associated condensates.  Interestingly, the back reaction to the metric with $\phi_2^I=0$ perturbs only the metric $\tilde g_{\mu\nu}$ but leaves $a(r)$ alone, implying that the UV and IR physics is the same.  Our cosmological solution is the de Sitter analogue when $V(\Phi)$ has a stationary point with a positive cosmological constant.

{\bf Conclusion and further discussion}: We constructed the cosmological model describing isotropic expansion of inhomogeneous universe.  The energy-momentum tensor $T^{\rm s}_{\mu\nu}$ that can cause inhomogeneity in space may not affect the isotropic expansion factor $a(t)$ provided that it satisfies (\ref{menergy}).  Such an energy-momentum tensor may not be exotic.  In fact, we showed that any $\sigma$-model could have this property.  The isotropic scaling factor $a(t)$ is driven by the energy-momentum tensor $T^{\rm t}_{\mu\nu}$ of perfect fluid, exactly the same as in the standard FLRW model.

The equations of motion of our model are also consistent with cosmological inflation. We demonstrated that an inhomogeneous universe could undergo an isotropic inflation.  The initial inhomogeneities are expected to be washed out by the inflation for large e-foldings.  However, some extreme large-scale inhomogeneity such as the dipole anisotropy may survive. They can also play an important role for some inflationary models with a moderate total e-folding number.  In some aspects, the energy-momentum tensor $T^{\rm s}_{\mu\nu}$ resembles that of dark matter.  They do not interact with other matter, but affect their motions gravitationally by creating the space curvature. However, they expand uniformly and do not contribute to $a(t)$ as dark matter does.  Our classical initial inhomogeneity resembles the primordial quantum fluctuation that was fixed in the co-moving frame. It is of interest to investigate whether such inhomogeneity has the effect of congregating baryonic matter in our universe, since this may give rise to large scale structures that are beyond the linear primordial quantum fluctuation.

We constructed some exact inhomogeneous spatial solutions with spherical and axial symmetries, including wormholes. For the spherically-symmetric system, we showed that the ans\"atze for cosmology and black hole formation were of the same class. We obtained some new exact black hole formation in cosmological gravity coupled to a free scalar from the simpler cosmological solution by the coordinate transformation.

The fact that some inhomogeneous spaces can expand isotropically is of great interest.  It does not require exotic matter and may have implication in cosmology; it is worth further investigation.

\section*{Acknowledgement}

We are grateful to Yi-Fu Cai, Qing-Guo Huang, Ming-Zhe Li, Jiro Soda, Robert Wald, Zhao-Long Wang  and Xuefeng Zhang for useful discussions.  The work is supported in part by the NSFC grants 11175269 and 11235003.

\end{document}